\documentclass[amsmath]{JHEP3}
\usepackage{graphicx}
\usepackage{epsfig}
\usepackage{amsmath}
\usepackage{bbm}
\newcommand{\der}[2]{\ensuremath{\frac{\partial #1}{\partial #2}}}
\title{Dynamical overlap fermions, results with hybrid Monte-Carlo algorithm}
\author{
Z. Fodor$^{a,b}$, S.D. Katz$^{a,c}$\thanks{On leave from Institute for Theoretical
Physics, E\"otv\"os University,
Budapest, Hungary.} and K.K. Szab\'o$^{a,b}$\\
\it $^a$Department of Physics, University of Wuppertal,
Germany\\
\it $^b$Institute for Theoretical Physics, E\"otv\"os University,
Budapest, Hungary\\
\it $^c$Deutsches Elektronen-Synchrotron DESY,
Hamburg, Germany}

\date{\today}
\maketitle
\abstract{
We present first, exploratory results of a hybrid Monte-Carlo algorithm
for dynamical, $n_f$=2, four-dimensional QCD with overlap fermions. 
As expected, the computational requirements 
are typically two orders of magnitude larger for the dynamical overlap formalism 
than for the more conventional (Wilson or staggered) formulations.
}

\begin{document}
\vspace*{-8.0cm}
\noindent
\hfill \mbox{WUB 03-08}
\hfill \mbox{ITP-Budapest 596}
\hfill \mbox{DESY 03-113}

\vspace*{7.8cm}

\vspace*{0.2cm}

\section{Introduction}

Fermionic operators (D) satisfying the
Ginsparg-Wilson relation \cite{Ginsparg:1981bj}
\begin{equation}\label{ginsparg-wilson}
\gamma_5 D+ D \gamma_5=\frac{1}{m_0} D \gamma_5 D
\end{equation}                                                         
allow to solve the chirality problem of 
four-dimensional QCD at non-vanishing lattice spacing 
\cite{Neuberger:1997fp}---
\cite{Neuberger:1998my}(for 
different reviews 
at recent lattice conferences see e.g.
\cite{Niedermayer:1998bi}---\cite{Giusti:2002rx}). 

Clearly, it would be very useful to exploit these new developments
in numerical studies of QCD. In the past years 
several groups have calculated already e.g. the quenched hadron spectrum
and light quark masses with better and better accuracies
(for recent results see the overlap formulation 
\cite{Hernandez:2001yn}---\cite{Chiu:2003iw}
and a related work with the perfect and chirally improved actions
\cite{Gattringer:2003qx}).

Due to limitations in computational resources no result is available
for dynamical, four-dimensional QCD with Ginsparg-Wilson fermions.
Some exploratory studies were carried out in the Schwinger model
and suggestions were made, which could help the  
four-dimensional full 
QCD case 
\cite{Narayanan:1995sv}---\cite{Giusti:2002sm}.

In this letter we present exploratory tests using
dynamical, four-dimensional QCD with Ginsparg-Wilson fermions.
We start with the Zolotarev optimal
rational polynomial approximation \cite{vandenEshof:2002ms}. 
The partial fraction expansion of the rational polynoms leads to
a particularly simple expression for the fermionic force of 
the hybrid Monte-Carlo.  In addition to the usual
fermion matrix inversion we have another inversion due to 
the inverse in the partional fraction expansion. These nested inversions
are very CPU consuming. By projecting out the smallest eigenmodes
in the inner loop a significant speed up could be reached.

As we emphasized, our results are exploratory. In addition
they are obtained on absurdly small lattices. As usual, direct physical 
interpretation will be possible only after studying larger lattices and
approaching the continuum limit.
Nevertheless, these first results
can be used as references in order to cross check future studies.

\section{Hybrid Monte-Carlo for QCD with overlap fermions}

First we fix our notations. The massless Neuberger-Dirac operator (or overlap 
operator) $D$ can be written as 
\begin{equation}\label{overlap}
D=m_0 [1+\gamma_5 {\rm sgn}(H_W)],
\end{equation}
This $D$ operator satisfies eq. (\ref{ginsparg-wilson}).
$H_W$ is the hermitian Dirac operator, $H_W=\gamma_5 D_W$,
which is built from the massive Wilson-Dirac operator, $D_W$, 
defined by
\begin{eqnarray}\label{wilson}
[D_W]_{xy}=\delta_{xy}-\kappa_W \sum_{\mu}
\left\{
U_{\mu}(x)(1+\gamma_\mu)\delta_{x,y-\mu}+
U_{\mu}^{\dagger}(y)(1-\gamma_\mu)\delta_{y,x+\mu}
\right\},
\nonumber
\end{eqnarray}
where $\kappa_W$ is related to $m_0$ by $\kappa_W^{-1}={8-m_0/2}$. 
The mass is introduced in the overlap operator by
\begin{equation}
D(m)=(1-m)D+m.
\end{equation}

In the sign function of eq. (\ref{overlap}) one uses 
${\rm sgn} (H_W)=H_W/\sqrt{H_W^2}$. 
The $n^{\rm th}$ order Zolotarev optimal rational approximation 
for $1/\sqrt{x}$ in some interval $[1,x_{\rm max}]$ can be expressed 
by elliptic functions (see e.g. \cite{Chiu:2002eh}).
For most of the purposes a more transparent
suboptimal choice $\epsilon(x)$ with $\epsilon(1)=1$ 
is sufficient\footnote{In
our test we use a 20th order approximation with e.g. $x_{\rm max}$=$10^{11}$.
It is easy to check that this choice corresponds to  
a relative accuracy of ${\cal O}(10^{-5})$. The difference 
between the optimal 
and suboptimal choices is one order of magnitude smaller.} 

\begin{equation}\label{zolotarev}
{\rm sgn}(x)\approx\epsilon(x)=
{
x\prod_{l=1}^{n}{(x^2+c_{2l})/(1+c_{2l})
\over
(x^2+c_{2l-1})/(1+c_{2l-1})}
},
\end{equation}
where
\begin{equation}\label{coeff}
c_l=
{
{\rm sn}^2(lK/(2n+1);\kappa)
\over
1-{\rm sn}^2(lK/(2n+1);\kappa)
},
\qquad\qquad \kappa=\sqrt{1-1/x_{max}},
\end{equation}
the Jacobian elliptic function ${\rm sn}(u,\kappa)=\eta$ is defined
by the elliptic integral
\begin{equation}
u(\eta)=\int_0^\eta
{
dt
\over
\sqrt{(1-t^2)(1-\kappa^2t^2)}
},
\end{equation}
and $K=u(1)$ is the complete elliptic integral. 
A particularly useful form of eq. (\ref{zolotarev}) and
an approximation for the sign function is given by 
partional fractioning
\begin{equation}\label{parcfrac}
 \epsilon(x)
=x(x^2+c_{2n})\sum_{l=1}^n {b_l \over x^2+c_{2l-1}},
\end{equation}
where the $b_l$ parameters are expressed by
the $c_l$ coefficients of eq. (\ref{coeff}).
In the rest of the paper we use the approximate $\epsilon(x)$
instead of the sign function in the Dirac operator
\begin{equation}
D=m_0 [1+\gamma_5 \epsilon(h_W)],
\end{equation}
where  we normalize  $H_W$
by its smallest eigenvalue: $h_W=H_W/|\lambda_{\rm min}|$. This choice
ensures that all the eigenvalues of $h_W^2$ are within the interval $[1,x_{{\rm max}}]$
where $x_{{\rm max}}$ is taken to be larger than the 
condition number of $H_W$.

We follow the standard procedure to implement 
the fermionic operator of eq. (\ref{overlap})
into a hybrid Monte-Carlo \cite{Duane:de} QCD algorithm. 
We analyze two flavours, thus $D^\dagger D$ is used as the Dirac  
operator. The fermionic
determinant for these two flavours can be given by introducing
the pseudofermionic fields
\begin{equation}
\det(D^\dagger D)=\int d\phi^\dagger d\phi \exp (-S_p) 
\qquad\qquad {\rm with}\qquad\qquad S_p=\phi^\dagger (D^\dagger D)^{-1} \phi.
\end{equation}
As usual, the integral is calculated stochastically by 
generating Gauss distributed $\phi$ pse\-u\-do\-fermions.
The contribution
of the pseudofermions to the force has the usual form 
\begin{equation}
{dS_p \over dU}=-\psi^\dagger {d D^\dagger D\over dU} \psi 
\qquad\qquad {\rm with}\qquad\qquad \psi=(D^\dagger D)^{-1} \phi .
\end{equation} 

Compared to the 
hybrid Monte-Carlo with Wilson fermions the new feature is
the somewhat more complicated structure of the
overlap operator. This complication 
is threefold. \hfill\break
a. First of all, it appears in the inversion of the fermion operator.
\hfill\break
b. Secondly, the complication is present in the structure of 
the derivative term in the fermionic force.
\hfill\break
c. Thirdly, the fermion force has Dirac-delta type singularities.

{\it ad a.} The inversion of the fermion operator 
$\psi=(D^\dagger D)^{-1}\phi$ 
is done by $n_o$ conjugate gradient steps (``outer inversion''). Note, 
however, that each step in this procedure needs the calculation of 
$(D^\dagger D)\phi$. The operator $D$ contains 
$\epsilon(h_W)$, which is given by the  partional fraction expansion 
(see eq. \ref{parcfrac}). Thus, at each ``outer'' conjugate
gradient step one needs $n$ different
inversions. Fortunately, these inversions differ
only by a constant term $c_{2l-1}$ ($l=1,...,n$). 
It means, that this ``inner inversion''
can be done by one multi-shift conjugate gradient \cite{Frommer:1995ik} 
procedure in $n_i$ steps, and one is not forced to carry out $n$
different conjugate gradient inversions. This nested conjugate
gradient procedure needs all together $n_o\cdot n_i$ matrix-vector
multiplications. 
It is already well known from the quenched analysis, 
that the number of steps in the inner inversion
can be significantly reduced by projecting out the smallest
eigenmodes and performing the conjugate  gradient steps
only in the orthogonal subspace. 

{\it ad b.} In the fermionic force the derivative with respect 
to the link variable $U$ can be 
straightforwardly calculated from the partial fraction
expansion eq. (\ref{parcfrac}). The term coming from the $\epsilon$
function reads
\begin{multline}
\psi^\dagger \gamma_5 {d \epsilon(h_W) \over dU}\psi
=\psi^\dagger \gamma_5 \sum_{l=1}^n 
\left\{ \frac{dh_W}{dU}(h_W^2+c_{2n})+\right. \\
\left. +(c_{2l-1}-c_{2n}) \frac{h_W}{h_W^2+c_{2l-1}}(h_W\frac{dh_W}{dU} +
\frac{dh_W}{dU} h_W) 
\right\} b_l\psi_l, 
\end{multline}
where the definition 
\begin{equation}
\psi_l=(h_W^2+c_{2l-1})^{-1}\psi
\end{equation}
was introduced. In order to calculate the force one
has to determine $\psi_l$. The above
inversion for the force is done by a multi-shift conjugate gradient
process in additional $n_f$ steps. Note, however,
that this inversion increases the computational
effort only marginally. 
Having obtained $\psi$ in ($n_o$$\cdot$$n_i$) multiplication steps,
$\psi_l$ can be obtained just by inverting $h_W^2+c_{2l-1}$. All together
the determination of $\psi_l$ needs ($n_o$$\cdot$$n_i$+$n_f$) 
matrix-vector multiplications.  

{\it ad c.} Since the fermionic force is the derivative of a 
non-analytic function, we expect non-trivial behaviour near these
non-analyticities. This feature is already present in a classical 
one-dimensional motion of a point-particle in a step
function potential. A finite stepsize integration of the equation of motion will not notice the step in the
potential or the Dirac-delta in the force. 
As a consequence the action has a large change which might lead to 
bad acceptance rate in the Monte-Carlo simulations. One can improve
on this situation.
During the integration one should check whether the particle moved from
one side to the other one of the step function. If it is necessary, 
one corrects its momentum and position.
This correction has to be done also in the case of the overlap fermion.
The microcanonical energy,
\begin{equation}
\mathcal{H}=
\frac{1}{2}\langle H,H\rangle+S_{\rm gauge}[U]+S_{\rm p}[U,\phi]=\frac{1}{2}\langle H,H\rangle
+S[U,\phi]
\label{eq:ener}
\end{equation}
has a step function type non-analyticity on the 
the zero-eigenvalue surfaces of the $H_W$
operator in the space of link variables coming from the pseudofermion action. 
In eqn. (\ref{eq:ener}) the 
$\langle A, B \rangle=-\sum_{x,\mu}{\rm tr}(A_{x,\mu}B_{x,\mu})$ scalar product
and $H$ anti-hermitian gauge momenta were introduced.
When the microcanonical trajectory reaches one of these surfaces, 
we expect either reflection or refraction.
If the momentum component, 
orthogonal to the zero-eigenvalue surface, 
is large enough to compensate the change of the action
between the two sides of the singularity ($\Delta S$)
then refraction should happen,
otherwise the trajectory should reflect off the singularity surface.
Other components of the momenta are unaffected.  
\begin{table}
\begin{center}
\begin{tabular}{|c|c|c|}
\hline
& When & New momenta \\
\hline
Refraction & $\langle N,H\rangle ^2 > 2\Delta S$ &
$H-N\langle N, H \rangle + N \sqrt{\langle N,H \rangle ^2-2\Delta S}$ \\
\hline
Reflection & $\langle N,H\rangle ^2 < 2\Delta S$ &
$H-2N\langle N,H \rangle$ \\
\hline
\end{tabular}
\end{center}
\caption{\label{tab:frref} Refraction and reflection can happen to the system when approaches a zero-eigenvalue
surface of $H_W$. The conditions and the new momenta are indicated. $H$ is
the momentum before the refraction/reflection.}
\end{table} 
The anti-hermitian normal vector ($N$)
of the zero-eigenvalue surface 
can be expressed with the help of the gauge derivative 
(in our shorthand notation $D\lambda$) as 
\[N=\left.\frac{D\lambda} {\sqrt{ \langle D\lambda,
D\lambda \rangle}}\right|_{\lambda=0},\] 
where $D\lambda = \langle \lambda
| D H_W | \lambda \rangle$. Table \ref{tab:frref} summarizes the 
conditions of refraction and reflection and the new momenta.

We have to modify the standard leap-frog integration of the equations of
motion in order to take into account reflection and refraction. This can be 
done in the following way. The standard leap-frog consists of three steps:
an update of the links with stepsize $\tau/2$, an update of the 
momenta with $\tau$ and finally another update of the links, using the new 
momenta, again with $\tau/2$, where $\tau$ is the stepsize of the integration.
The system can only reach the zero-eigenvalue surface during the update of the
links. We have to identify the step in which this happens. This can be done
by predicting the change of the lowest eigenvalues during 
one elementary link update using the derivatives $D\lambda$. 
At the stepsizes we used,
this procedure turned out to be very reliable. 
After identifying the step in which the zero eigenvalue surface is reached, 
we have to replace it with the following three steps:
\begin{enumerate}
\item Update the links with $\tau_1$, so that we reach exactly the zero-eigenvalue
surface. $\tau_1$ can be determined with the help of $D\lambda$.

\item Modify the momenta according to Table \ref{tab:frref}. 

\item Update the links using the new momenta, with stepsize $\tau/2-\tau_1$.
\end{enumerate}
This procedure is trivially reversible and it also preserves
the integration measure (see Appendix).

Including reflection and refraction into the update is a crucial point
in the simulations. If the system does not notice the step in the 
action due to the finite stepsize integration, there
will be huge jumps in the energy leading to a very bad acceptance ratio 
in the final accept/reject step. 

\section{Numerical tests}

As it was discussed by many authors (and we also illustrated above) 
the dynamical hybrid Monte-Carlo for QCD with overlap fermions 
is computationally extremely intensive due to the nested inversion.
${\cal O}(100)$ conjugate gradient steps
is usually enough for Wilson or staggered fermions. 
In the overlap formalism one is confronted
with ${\cal O}(100^2)$ matrix-vector multiplications. 
Therefore, with a straightforward hybrid Monte-Carlo and with 
present medium-size machines only absurdly
small lattices can be studied. Nevertheless, these
studies can show the feasibility of the algorithm
and can be used for cross-checking future results.

We studied our hybrid Monte-Carlo on 
$V=2^4,4^4$ and on $4\cdot 6^3$ lattices with $m_0=1.6$, with 
mass parameters
$m=0.1,0.2$ and $\beta$ between $5.3$ and $6.1$. The length
of our trajectories were 1. We used $\Delta\tau=0.01$---$0.05$
as time-steps for the molecular dynamical evolution. 
The Metropolis accept/reject step at the end of the
trajectories resulted in a 30-80\% acceptance rate.

We used a 20th order rational polynomial approximation.
This choice gives the sign function with  
a relative accuracy of ${\cal O}(10^{-5})$. Note, that changing
the order of the approximation from 10 to 20 increased 
the computational effort only by 20\%. 

In order to
accelerate the inner inversion we projected out the
eigenmodes with the smallest eigenvalues. The inversion was 
then performed in the orthogonal subspace. We studied the
computational requirements as a function of the number of the
projected eigenmodes. The projections were done by the 
ARPACK code. The optimum was found around
20 eigenmodes. The projection leads to an important observation.
The operator $H_W$ might have rather small eigenvalues e.g. 
${\cal O}(10^{-6})$. 
In order to project out eigenmodes one has to 
solve eigenvalue equations. In these equations
the sum of ${\cal O}(1)$ numbers should result in     
${\cal O}(10^{-6})$. This is clearly beyond the
accessible region of 32-bit arithmetics. Therefore, we used
64-bit precision. 

In addition to the standard consistency tests
(reversibility of the trajectories and 
$\Delta\tau^2$ scaling of the action)
we performed a brute force approach on $2^2$ and
$4^4$ lattices. We generated quenched
configurations, then we explicitely calculated the
determinants of $m_0 [1+\gamma_5 \epsilon(h_W)]$.
These determinants were used in an additional 
Metropolis accept/reject step.
The hybrid Monte-Carlo results agree completely with 
those of the brute force approach.

Table~\ref{data_set}
gives informations on our run parameters and
test results.

\renewcommand{\arraystretch}{1.2}
\begin{table}[h!]
\begin{center}
\begin{tabular}{|c|c|c|c|}
\hline
V    & ($\beta$,m) & number of trajectories   & Plaquette   \\
\hline
$2^4$           & $(5.6,0.2)$   & $900$   &  $3.44(1)$  \\
\hline
$4^4$           & $(5.4,0.2)$   & $1200$    & $2.572(4)$   \\
\hline
$4\cdot 6^3$           & $(5.78,0.1)$   & $400$   & $3.22(1)$   \\
\hline
\end{tabular}
\end{center}
\caption{\label{data_set}
Expectation values of the plaquette variable for different volumes
and parameters.}
\end{table}
\begin{figure}[h]
\begin{center}
\includegraphics*[width=7.0cm,bb=20 170 570 700]{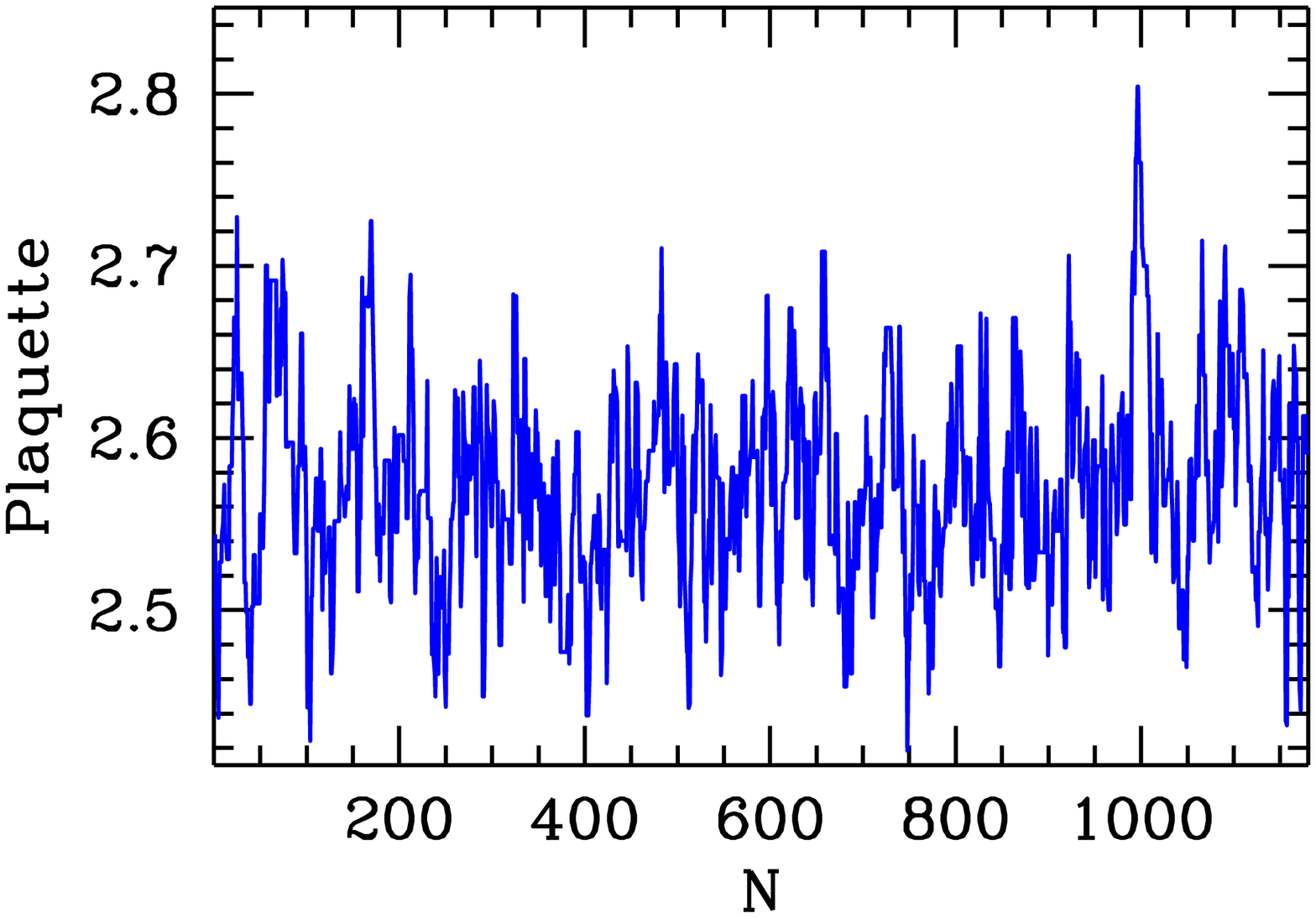}
\includegraphics*[width=7.0cm,bb=20 170 570 700]{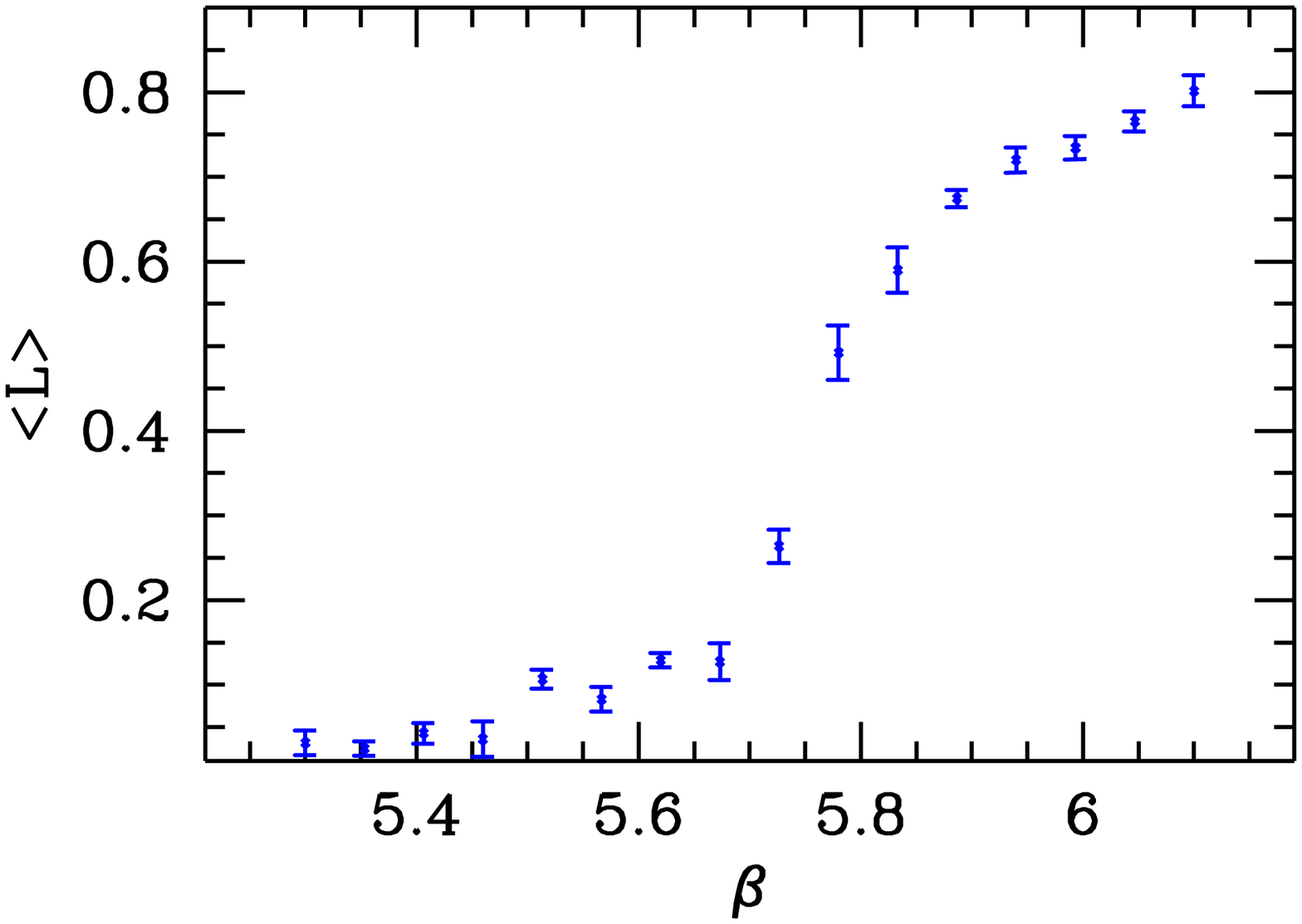}
\caption{\label{time_history}
The time history of the plaquette (left panel) 
on a $4^4$ lattice 
at $m=0.2$ and $\beta=5.4$. 
The $\beta$ dependence (right panel)
of the Polyakov-loop on $4\cdot 6^3$ lattices at $m=0.1$.
(Quenched results suggest that the pion mass could be around 200--250~MeV
for our parameter choice at $\beta$=5.7)}
\end{center}
\end{figure}

The left panel of Figure~\ref{time_history} shows the time
history of the plaquette variable in one of our $4^4$ runs. 
We also observed 
changes between
topological sectors. 
The ratio of topologically non-trivial and trivial gauge
configurations was found to be about the same as in the brute force approach 
(it is around or below the percent level).  
On the right panel we present
the $\beta$ dependence of the Polyakov-loop. These runs were obtained on
$4\cdot 6^3$ lattices
\cite{Katz:2003up}. One can see a sharp increase around $\beta \simeq 5.7$.  
As usual, before drawing any physical interpretation
one should proceed to the continuum limit (see also 
Ref. \cite{Golterman:2003qe}).

\section{Conclusions}

In this paper we described a hybrid Monte-Carlo algorithm
for dynamical, $n_f$=2, four-dimensional QCD with overlap fermions. 
We used a modified version of the MILC collaboration's code.
We started with the Zolotarev optimal rational polynomial
approximation to numerically implement the sign function.
Using the partional fraction expansion of the
rational polynomial resulted in a particularly
simple form for the fermionic force. The inversions
due to the fermion operator (``outer inversion'')
and due to the rational polynomial denominator (``inner inversion'')
were done successively by conjugate gradient processes. It was 
possible to significantly accelerate the inner inversion    
by projecting out the lowest eigenmodes and to use a multi-shift
solver for the different terms in the partional fraction 
expansion. 

We extended the standard leap-frog integration of the trajectories by
including the refraction and reflection on the zero-eigenvalue surfaces
of the Wilson-Dirac operator. The inclusion of these effects
increased the acceptance rate of the algorithm significantly.

We compared our hybrid Monte-Carlo results with those
obtained by a brute force approach (quenched configurations with 
Metropolis accept/reject steps
for the exactly calculated overlap determinant). 
A complete agreement was found
for $2^4$ and $4^4$ lattices.

When writing up this paper an independent hybrid Monte-Carlo  
was written \cite{Lippert:2003xxx}. 
We cross-checked the results for $4^4$ lattices 
and a complete agreement was found.

\vspace{0.5cm}  
\noindent
{\bf Acknowledgments:\\} 
Our tests were done by a modified version of 
the MILC Collaboration's public code\\ 
(see http://physics.indiana.edu/\~{}sg/milc.html).
Numerical tests were carried out on the PC-cluster (163 Intel-P4 nodes)
at the E\"otv\"os University of Budapest, Hungary.
Useful discussions and comments on the manuscript from
Th. Lippert, I. Montvay and H. Neuberger are acknowledged. Special
thanks go to T. Kov\'acs for his comments and for
bringing two bugs of the open ARPACK code into our attention.
This work was partially supported by Hungarian Scientific
grants, OTKA-T34980/\-T29803/\-T37615/\-M37071/\-OMFB1548/\-OMMU-708.

\vspace*{0.5cm}
\noindent
{\large \bf Appendix}
\vspace*{0.3cm}

In this appendix we examine the area-preserving property of the modified leap-frog procedure 
described in the text. 

Let us start with an example of the 
$N$-dimensional Euclidean coordinate space which shows the basic idea of the
proof in a transparent way.

We solve the equations of motion with a finite stepsize integration
of the following Hamiltonian:
\begin{equation*}
{\cal H} = \frac{1}{2}p_a p_a +S\left({\rm sgn}M(q) \right),
\end{equation*}
where $q_a, p_a$ $(a=1\dots N)$ are the coordinates and the momenta. 
$M$ depends only on the coordinates and the action $S$
is a smooth function (note that $q_a$, $M$ and $S$ are analogous
to the links, the fermion matrix and the fermionic action, respectively). 
The standard leap-frog algorithm can be effectively 
applied to this system, as long
as the trajectories do not cross the 
zero-eigenvalue surface of $M$ ($\lambda(q)=0$, where $\lambda(q)$ is the 
eigenvalue with smallest magnitude\footnote{We do not deal with the 
possibility of degenerate zero eigenvalues which appears only
on a zero measure subset of the zero-eigenvalue surface.}).
                          
We have to modify the leap-frog algorithm, 
when the coordinates  
reach the zero-eigenvalue surface. 
Instead of the original leap-frog update of the coordinates, where
the constant $p_a$ momenta are used for the time $\tau/2$, 
we first update the coordinates with $p_a$ until the surface, then we change 
the momentum to $p'_a$, which is used to evolve $q_a$ for the remaining 
time. In case of refraction one has the following 
phase space transformation:
\begin{eqnarray}
\label{eq:transform}
q'=q+\tau_1 p+(\tau/2-\tau_1)p' \\
p'=p-n(np)+n(np') \nonumber
\end{eqnarray}
where $n$ is the normalvector of the surface, $\Delta S$ is the potential jump along the surface, and  
$(np')^2=(np)^2-2\Delta S$. $\tau_1$ is the time required to reach the surface with the incoming momenta $p$.  
The transformation for  reflection is given by
\begin{eqnarray}
\label{eq:transform1}
q'=q+\tau_1 p+(\tau/2-\tau_1)p' \\
p'=p-2n(np) \nonumber
\end{eqnarray}
In the following we will not deal with this case (one can obtain 
the Jacobian of reflection by simply setting
$(np')=-(np)$ in the Jacobian of refraction). 

First let us concentrate on the $q,p$ dependence of $\tau_1$. 
$\tau_1(q,p)$ is determined from the  condition
\(
\lambda(q+\tau_1(q,p)p)=0
\).
One obtains the partial derivatives of $\tau_1$ with respect to $q,p$
by expanding this zero-eigenvalue 
condition to first order in $\delta q$ or $\delta p$.
First take the $\delta q$ variation:
\begin{align*}
\lambda(q_a+\tau_1p_a+\delta q_a+\der{\tau_1}{q_b}\delta q_b p_a)=\lambda(q+\tau_1p)+\left.\der{\lambda}{q_a}\right|_{q+\tau_1p}
(\delta_{ab}+\der{\tau_1}{q_b}p_a)\delta
q_b=0
\end{align*}
Since the normalvector is just 
\[
n_a=\left.\der{\lambda}{q_a}\right|_{q+\tau_1p}/||\der{\lambda}{q}||, 
\]
we have for the partial derivative of $\tau_1$ with respect to $q$:
\[
\der{\tau_1}{q_a}=-\frac{n_a}{(np)}.
\]
Similarly one gets for the partial derivative with respect to $p$:
\[
\der{\tau_1}{p_a}=-\tau_1\frac{n_a}{(np)}.
\]

There is an important identity between the $q$ and $p$ derivatives of a function, which depends 
only on
$q+\tau_1(q,p)p$. (Two examples are $n$ and $\Delta S$.)
Let us evaluate $p$ and $q$ derivatives of an arbitrary 
$g(q+\tau_1(q,p)p)$ function:
\begin{align*}
\der{g}{q_a}=\left.\der{g}{q_b}\right|_{q+\tau_1p}(\delta_{ab}+\der{\tau_1}{q_a}p_b)=\left.\der{g}{q_b}\right|_{q+\tau_1p}
(\delta_{ab}-\frac{n_ap_b}{(np)}),
\\ 
\der{g}{p_a}=\left.\der{g}{q_b}\right|_{q+e_1p}(\tau_1\delta_{ab}+\der{\tau_1}{p_a}p_b)=\left.\der{g}{q_b}\right|_{q+\tau_1p}
(\delta_{ab}-\frac{n_ap_b}{(np)})\tau_1,
\end{align*}
which gives
\begin{eqnarray} 
\der{g}{p_a}=\tau_1\der{g}{q_a}.
\label{eq:pq}
\end{eqnarray}

Now we can consider the four different partial derivatives required for the Jacobian:
\[
J=
\begin{pmatrix}
\der{q'}{q} & \der{q'}{p} \\
\der{p'}{q} & \der{p'}{p} \\
\end{pmatrix},
\]
whose determinant gives the change in the Euclidean measure $d^Nqd^Np$ due to the given
phase space transformation.
Introducing 
\[
Z_{ab}\equiv \der{p'_a}{q_b}.
\]
one incorporates all terms which arise from the $q$ 
dependence of the normalvector and $\Delta S$. 
In case of a straight wall with constant potential jump this matrix vanishes. 
(Clearly, for QCD with
overlap fermions these objects are very hard to calculate;
they usually require the 
diagonalization of the whole $H_W$ matrix ).
Using (\ref{eq:pq}) the other three components of 
$J$ can also be expressed with the help of $Z$. 
Denoting
\begin{align*}
x\equiv\frac{(np')}{(np)}-1, &&  y\equiv\frac{(np)}{(np')}-1 \\
\end{align*}
and
\begin{equation*}
P_{ab}\equiv\delta_{ab}+xn_an_b,
\end{equation*}
we have 
\begin{align*}
P^{-1}_{ab}=\delta_{ab}+yn_an_b.
\end{align*}
In terms of the $P$, $P^{-1}$ and $Z$ matrices the 
Jacobian is very simple.
We can split it into 2 parts: the first part is a matrix with determinant 
one and all $Z$ factors are in the second term:
\[
J=
\begin{pmatrix}
P & \tau_1P+(\tau/2-\tau_1)P^{-1} \\
0 & P^{-1}
\end{pmatrix}+
\begin{pmatrix}
(\tau/2-\tau_1)Z & (\tau/2-\tau_1)\tau_1Z \\
Z & \tau_1 Z
\end{pmatrix}.
\]
Let us introduce $J'$ as the product of $J$ and the inverse of its first term:
\begin{align*}
J'
=\begin{pmatrix}
1 & 0 \\
0 & 1
\end{pmatrix}\otimes \mathbbm{1}
+E\otimes \tau_1PZ,
\end{align*}
where $E$ is defined as
\[
E=\begin{pmatrix}
-1 & -\tau_1 \\
1/\tau_1 &  1
\end{pmatrix}.
\]
$E$ has an eigenvector $v_1 \propto (\tau_1,-1)$ with zero eigenvalue. 
The $v_2\propto (1,\tau_1)$ vector is orthogonal
to $v_1$ and
has the property to give zero in the product $v_2^TEv_2=0$. In the orthonormal 
basis given by $v_1$ and
$v_2$ $J'$ has the form:
\begin{align*}
J'
=
\begin{pmatrix}
1 & 0 \\
0 & 1
\end{pmatrix}
\otimes \mathbbm{1}
+ 
\begin{pmatrix}
0 & v_1^TEv_2 \\
0 & 0
\end{pmatrix}
\otimes \tau_1PZ,
\end{align*}
thus $\det J'=1$. Since $J$ and $J'$ differs only in a matrix with determinant one, we arrive
\[
\det J=1,
\]
thus the transformations (\ref{eq:transform}, \ref{eq:transform1}) preserve 
the integration measure. 

The proof for the $SU(3)$ case was carried out in a completely analogous
way. The only difference was the appearance of factors
associated with the group structure of $SU(3)$ which all canceled out in 
the final result. Thus, we conclude that the suggested modification of the
leap-frog conserves the integration measure. 


\end{document}